\documentclass[11pt]{article}
\usepackage[a4paper,margin=2.4cm]{geometry}
\usepackage[T1]{fontenc}
\usepackage[utf8]{inputenc}
\usepackage{lmodern}
\usepackage{amsmath,amssymb}
\usepackage{graphicx}
\usepackage{booktabs}
\usepackage{array}
\usepackage{microtype}
\usepackage{xcolor}
\usepackage[hidelinks]{hyperref}
\usepackage{url}
\usepackage{xspace}
\usepackage[section]{placeins}
\usepackage{caption}
\graphicspath{{figures/}}
\newcommand{\HU}{\ensuremath{\,\mathrm{HU}}\xspace}

\title{RIQE: a NIQE-style reference model for Computed Tomography}

\author{Fabio Mattiussi\\
\href{mailto:fabio.mattiussi@eoc.ch}{\texttt{fabio.mattiussi@eoc.ch}}}

\date{}

\begin{document}
\maketitle

\begin{abstract}
The Natural Image Quality Evaluator (NIQE) scores an image by its statistical distance from a
model fitted on pristine images, and its distributed model is fitted on photographs. We
release the Radiology Image Quality Evaluator (RIQE), a NIQE-style model fitted on 3{,}792
full-dose slices from 158 patients of the public LDCT-and-Projection-data collection, with a
declared intensity mapping, a manifest of every slice and a script that reproduces the fit.
On 40 held-out patients, RIQE ranks reduced-dose reconstructions, simulated by
projection-domain noise insertion, worse than the full-dose reconstruction of the same slice
in 240 of 240 chest and 230 of 240 abdominal pairs, and ranks images with 20\% more noise
worse than their source in 97.5--100\% of cases. Its preferences among filtered images,
however, do not follow lesion signal. With a 4\,mm, $+10\HU$ lesion inserted in noisy
abdominal slices, RIQE prefers bilateral filtering to the unfiltered image in every image up
to a 32\HU residual, at which 29\% of the lesion's matched-filter signal remains and its
detectability index falls from 0.51 to 0.33; it never prefers Gaussian smoothing, which at the
same 32\HU residual leaves 70\% of the signal and a detectability index of 0.48. Fitted on
photographs with the parameters published for NIQE, the same code ranks every simulated
reduced-dose abdominal image better than its full-dose counterpart. RIQE is suited to ranking
a degraded image against its source; under the conditions tested it should not be the sole
criterion for selecting, comparing or tuning denoisers.
\end{abstract}

\section{Introduction}

Blind, or no-reference, image quality assessment asks how far an image departs from the
statistical regularity expected of its class. NIQE~\cite{mittal2013niqe} builds that
expectation as a multivariate Gaussian over natural-scene-statistics (NSS) features
extracted from a corpus of pristine images, and scores a test image by a Mahalanobis-like
distance between its own fitted Gaussian and the corpus one. The construction is attractive
for CT because it needs neither a reference image nor human opinion scores, the two things a
low-dose or denoising pipeline cannot supply at inference time.

The obstacle is that the expectation is corpus-specific. NIQE's distributed model is fitted
on 125 natural photographs. A CT slice is not a photograph: its noise texture is set by the
reconstruction kernel and the dose, its intensities are attenuation values rather than
scene radiance, and its content has no analogue in outdoor photography. Scoring a CT image
against the photographic model yields a number that is well defined and reproducible, but
whose relation to CT image quality has to be shown rather than assumed.

Refitting the model on CT is easy; the model is 36 numbers and a $36\times36$ matrix. What
has not been published is evidence of how a CT-fitted model behaves on the two decisions it
would be used for. The first is \emph{ranking}: given an image and a degraded version of the
same anatomy, does the score order them correctly? The second is \emph{selection among
processed images}: given a noisy image and several filtered versions, does the score prefer
the ones that preserve diagnostic signal? The second is where a blind metric is most wanted,
and where a wrong preference does harm: aggressive filtering erases low-contrast lesions,
and a metric that improves while that happens selects the pipeline that removes the finding.
A review of medical image quality assessment~\cite{rodrigues2025review} reports a NIQE
variant adapted to medical imaging that ranked a \emph{noisy} ultrasound image as better than
the original.

\begin{figure}[tbp]
\centering
\includegraphics[width=\linewidth]{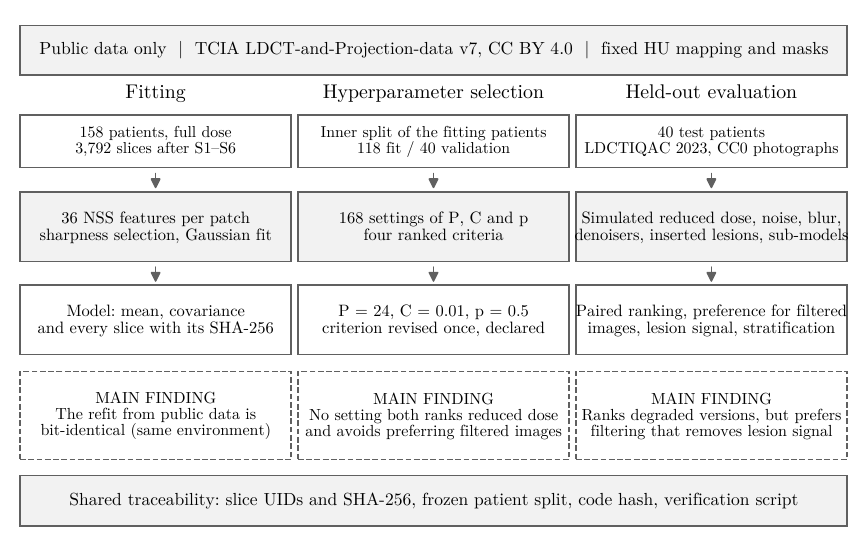}
\caption{Study design. Patients are split once, before any fitting. Hyperparameters are
selected on an inner split of the fitting patients (Figure~\ref{fig:tradeoff}), the
stratification analysis uses the fitting patients (Table~\ref{tab:strat}), and all other
results are computed on the 40 held-out test patients or on external data. Dashed boxes
summarise the main finding of each stage.}
\label{fig:design}
\end{figure}

This paper releases the model together with everything needed to rebuild it, and tests both
decisions. The CT-fitted model ranks degraded images correctly against their source. Its
preferences among filtered images do not follow the signal of inserted low-contrast lesions:
it prefers an edge-preserving filter at strengths that leave less than a third of the
matched-filter signal of a 4\,mm lesion, and does not prefer linear smoothing that leaves
more. Across the
hyperparameter settings tested, no single model met both our dose-ranking and filtering
criteria. Figure~\ref{fig:design} summarises the study.

\section{Related work}

\paragraph{NIQE.} NIQE~\cite{mittal2013niqe} computes mean-subtracted contrast-normalised
(MSCN) coefficients, fits a generalised Gaussian to their distribution and asymmetric
generalised Gaussians to the products of horizontally, vertically and diagonally adjacent
coefficients, following BRISQUE~\cite{mittal2012brisque}, at two scales, giving 36 features
per patch. At fitting time only patches whose local activity exceeds a fraction $p$ of the
image maximum are kept; at scoring time all patches are used. The original work reports stable
behaviour for patch sizes from $32\times32$ to $160\times160$ and for $p \in [0.6, 0.9]$ on
photographs. Section~\ref{sec:constraints} shows why neither range transfers directly to CT.

\paragraph{NIQE refitted on CT.} Gunawan et al.~\cite{gunawan2025} refit the pristine model
on standard-dose CT from the same collection we use and report an optimal patch size of
$16\times16$ with $p$ peaking at 0.32. The result is hard to reuse or check: the model
parameters and the code are not published; the conversion from Hounsfield units to
luminance, on which the features depend, is not stated; and hyperparameters are chosen to
maximise the fraction of images on which the deepest network is ranked the best denoiser, a
criterion that assumes the ordering the metric is meant to measure. We start from their
values and select against degradations of known direction instead.

\paragraph{Human-scored CT quality.} The LDCTIQAC~2023 challenge~\cite{lee2025ldctiqa}
released 1000 abdominal CT training images with the mean quality score of five
radiologists~\cite{ldctiqacdata}. The images derive from about 100 slices of seven patients,
each degraded in several ways, so they are far from 1000 independent observations. The
challenge report evaluates NIQE as a baseline on the challenge test set (Spearman
$-0.71$); those images and labels are not the ones we score, and the value is not directly
comparable with ours. Challenge methods are trained on the scores; RIQE is not. We use the
training data as an external check of perceived quality, which is distinct from diagnostic
performance.

\section{Methods}

\subsection{Data}

We used the TCIA collection LDCT-and-Projection-data~\cite{mccollough2020data,moen2021data},
version~7, accessed through the public NBIA REST interface~\cite{clark2013tcia}. In this
version the Chest and Liver subsets are released under CC~BY~4.0; the Head subset remains
under controlled access and was not used. We downloaded the reconstructed image series only:
299 series, 72{,}588 slices. The full-dose series fall into five protocol cells: Siemens
chest, B50f kernel, 1.5\,mm (50 patients); GE chest, STANDARD kernel, 1.25\,mm (41); GE
chest, CHST kernel, 1.25\,mm (8); Siemens abdomen, B30f, 5\,mm (50); and GE abdomen,
STANDARD, 5\,mm (50).

For the 100 Siemens patients the collection also provides a reduced-dose reconstruction of
the same examination. The data providers simulated it by inserting noise into the full-dose
projection data, with a validated noise-insertion method, to 10\% of the routine dose for
chest and 25\% for abdomen, and reconstructed the result with the same settings on the same
slice positions~\cite{moen2021data}. These reconstructions are not independent low-dose
acquisitions; we call them simulated reduced-dose images, or reduced dose for short. GE
patients have no reduced-dose series.

Two properties of the collection constrain the design. Kernel, slice thickness and anatomy
are not crossed: on each vendor, chest is always the sharper kernel and the thinner slice.
And the reconstruction diameter varies from 282 to 500\,mm, so pixel spacing spans 0.551 to
0.977\,mm while NSS features live on a pixel grid. We do not resample, since resampling is a
filter and would alter the statistics being measured; pixel spacing is instead tested as a
stratification axis.

\subsection{Intensity mapping}
\label{sec:mapping}

The conversion from Hounsfield units to luminance determines the features, so it is part of
the model specification:
\begin{equation}
L = 255\,\frac{\operatorname{clip}(\mathrm{HU}, -1000, 1000) + 1000}{2000},
\end{equation}
in 32-bit floating point without rounding. $-1000\HU$ is the physical value of air, and the
window retains the lung parenchyma that any soft-tissue window would destroy. $+1000\HU$ is
symmetric about water and clips 0.26\% of in-body pixels (measured on 144 slices across all
cells), which are dense cortical bone and metal. The 0--255 scale is kept because the MSCN
stabilising constant is defined on it (Section~\ref{sec:features}); quantisation to 8 bits is
not part of the algorithm and is not applied.

\subsection{Field-of-view and body masking}

GE images in this collection carry 55{,}772 pixels per slice (21.3\% of the area) at
$-3024\HU$ outside the reconstructed field of view; Siemens images carry none. Untreated, a
fifth of every GE image becomes a constant plane with zero local variance, and a between-vendor
difference appears that is a DICOM padding convention rather than acquisition physics. Every
patch must therefore lie entirely inside the field of view and at least 90\% inside a body
mask (HU $> -300$, $5\times5$ opening, largest connected component, holes filled). The
masks of an image are also used for all its degraded and filtered versions, and a
reduced-dose image is scored on the masks of the full-dose image at the same position, so
that a degradation cannot change the domain being scored.

\subsection{Features and model}
\label{sec:features}

Features follow the published description of NIQE~\cite{mittal2013niqe} and were implemented
from the paper without reference to any third-party code; we have not compared the whole
pipeline with the original implementation. MSCN coefficients are
$\hat I = (I - \mu)/(\sigma + C)$ with $\mu$ and $\sigma$ computed by a $7\times7$ Gaussian
window with standard deviation 1 (the literal reading of ``sampled out to 3 standard
deviations, $K = L = 3$''). Generalised and asymmetric generalised Gaussian parameters are
estimated by moment matching~\cite{sharifi1995,lasmar2009}; both estimators recover known
parameters from synthetic samples to within 1\%. The second scale uses the same Gaussian
kernel followed by 2:1 decimation. The model is the mean $\nu_1$ and covariance $\Sigma_1$ of
the selected patch features, and the score of an image with patch mean $\nu_2$ and covariance
$\Sigma_2$ is
\begin{equation}
D = \sqrt{(\nu_1 - \nu_2)^{\top}\left(\tfrac{\Sigma_1 + \Sigma_2}{2}\right)^{+}(\nu_1 - \nu_2)},
\label{eq:score}
\end{equation}
where $^{+}$ is the Moore--Penrose pseudo-inverse with singular values below $10^{-10}$ times
the largest set to zero. An image needs at least 72 valid patches, twice the feature
dimension, to have a defined score. A 36-dimensional covariance can be computed from 37
patches, but its estimate is unstable near that limit; 72 is a stability choice, not a
mathematical bound.

In NIQE $C = 1$, an absolute constant calibrated on 8-bit photographs in which local
$\sigma$ in structured regions is 10--40 code values. On our mapping the median local
$\sigma$ inside the body is 1.85 code values in Siemens abdomen and 7.60 in Siemens chest, so
$\sigma/(\sigma + 1)$ ranges from 0.65 to 0.88 across protocols: with $C = 1$ the divisive
normalisation is partial and protocol dependent, and absolute noise amplitude leaks into the
features. $C$ is therefore a declared hyperparameter.

\subsection{Pristine corpus}

Inclusion criteria are boolean functions of per-slice statistics with declared thresholds,
applied to a table so that they can be re-run and audited. We excluded the first and last
10\% of each series (9{,}572 slices), slices whose body covers less than 15\% or more than
85\% of the field of view (1{,}870), truncated slices (none beyond the previous criteria) and
slices in which more than $10^{-4}$ of in-body pixels exceed $2000\HU$ (736), and kept 24
evenly spaced slices per patient. The metal threshold was calibrated by visual inspection of
stratified samples: the $1500\HU$ threshold we first proposed is cortical bone, exceeded by
43\% of slices. A criterion intended to detect streak and motion artefacts from the local
$\sigma$ of air was falsified during calibration, since its extreme values identified large
patients with clean images and it moved in the wrong direction on bilateral hip prostheses;
it was disabled, and motion is therefore not detected directly. The resulting corpus has
4{,}752 slices from 198 patients. One patient was lost entirely, for a metallic foreign body
in the chest wall.

Patients, never slices, were split into a fitting set of 158 (with an inner 118/40 split for
hyperparameter selection) and a test set of 40, stratified by protocol cell and frozen before
any fitting.

\subsection{Test images}
\label{sec:testimages}

All degradations are applied in Hounsfield units before the intensity mapping.
Table~\ref{tab:flow} gives, for every experiment, what was available, the selection rule, the
exclusions and the patients and images analysed.
\begin{itemize}
\item \emph{Sampling.} Of the 24 kept slices of each test patient, six evenly spaced slices
(240 in all, 120 chest and 120 abdomen) receive the synthetic degradations and the filters;
the subsample keeps the image bank, 18{,}840 images, within our computing budget.
\item \emph{Reduced dose.} Every kept full-dose slice of the 20 Siemens test patients is
paired with the reduced-dose slice of the same patient at the same table position
($|\Delta z| \le 0.01$\,mm): 240 chest pairs from 10 patients and 240 abdominal pairs from 10.
Validation uses the same rule on the inner split (480 pairs).
\item \emph{Noise relative to native noise.} FBP-like correlated noise (white noise shaped by
a radial ramp with a Hann roll-off) added so that the median local standard deviation inside
the body grows by 5, 10, 20, 50 or 100\%. That median includes anatomical structure as well
as noise, and serves here as a proxy for native noise; the amount is calibrated in the units
of the local estimator, since for this noise the median local $\sigma$ is 0.637 times the
global one.
\item \emph{Noise at fixed levels.} FBP-like noise of 1 to 90\HU (14 levels) added to the full-dose image.
Three levels (5, 10 and 20\HU) are drawn twice with independent realisations; the median
absolute difference between the scores of the two realisations is the noise floor against
which a score change is compared.
\item \emph{Blur.} Gaussian, $\sigma$ from 0.5 to 3 pixels.
\item \emph{Filtering.} Gaussian, total variation~\cite{chambolle2004},
bilateral~\cite{tomasi1998}, non-local means~\cite{buades2005} and wavelet
BayesShrink~\cite{chang2000} from scikit-image~\cite{vanderwalt2014}. Native parameters are
not comparable across filters, so each filter is calibrated per image by bisection on its
strength until the residual $r = \operatorname{std}_M\!\big(f(I) - I\big)$, the standard
deviation inside the body mask $M$ of the change made by the filter $f$, reaches 2, 4, 8, 16,
32 or 64\HU. The residual measures how much a filter changes the image, not how much noise it
removes; it is the one strength scale common to the five filters. Of 10{,}800
calibrations, 25 (0.2\%) ended more than 3\% from the target, at most 22\%, mostly wavelet
shrinkage at 64\HU.
\end{itemize}

\subsection{Inserted lesions}
\label{sec:lesions}

Soft-edged discs of 4, 6 and 10\,mm at $+10$, $+15$ and $+25\HU$ are inserted at up to 16
automatically selected sites per size (at least 4, disjoint between sizes) in homogeneous
soft tissue: local mean 40--80\HU and local standard deviation below the image median. The
substrate is the central kept slice of each of the 20 abdominal test patients, of which
18 have room for the 4\,mm lesions. Enhanced liver, at about 130\HU in these
contrast-enhanced scans, falls outside the site range, so most sites lie in muscle. FBP-like
noise is added at the level that separates full-dose from reduced-dose Siemens abdomen. That
level is an operational estimate in the units of the local estimator,
$\sigma_n = (\sigma_{\text{low}}^2 - \sigma_{\text{full}}^2)^{1/2} = 18.4\HU$, where
$\sigma_{\text{low}}$ and $\sigma_{\text{full}}$ are the medians over slices of the median
local standard deviation inside the body. Since the local estimator returns 0.637 of the
global standard deviation of the synthetic noise, the noise is generated with a global
standard deviation of $18.4/0.637 = 28.9\HU$, so that its local standard deviation is
18.4\HU. The same noise is used on both vendors, since GE abdomen has no reduced-dose
series.

For site $s$ and noise realisation $k = 1, \dots, K$ ($K = 64$), with the background slice
$b$ fixed, the same realisation $n_k$ added with and without the lesion $l$, and the filter
$f$ applied at the strength calibrated once on the noisy slice without lesions,
\begin{equation}
A_{sk} = t_s^{\top} f(b + n_k + l), \qquad B_{sk} = t_s^{\top} f(b + n_k),
\end{equation}
where $t_s$ is the zero-mean, unit-norm disc template centred on the site. We report the
signal retained and the detectability index of the non-prewhitening matched filter,
\begin{equation}
R = \frac{\operatorname{mean}_{s,k}(A_{sk} - B_{sk})}{t^{\top} l}, \qquad
d' = \operatorname{mean}_s\,\frac{\operatorname{mean}_k(A_{sk} - B_{sk})}{\operatorname{sd}_k(B_{sk})},
\end{equation}
so that $R = 1$ without filtering. Retained signal is not retained detectability: a filter
also changes the variance in the denominator of $d'$, which is why both are reported. Holding
the background fixed removes anatomical variability from the ensemble, but not the
interaction between anatomy and a non-linear filter. The matched filter is a model observer
without eye filter or internal noise, not calibrated to human performance. The RIQE score is
that of the whole filtered noisy slice without lesions, so that the metric is compared with
lesions it cannot see.

\subsection{Photographic baselines}
\label{sec:baselines}

To separate the effect of the reference corpus from that of the code, the same code is also
fitted on 119 photographs released into the public domain or under CC0 (Wikimedia Commons;
luma by Rec.~601 on a 0--255 scale, at the distributed resolution, typically 1920 pixels
wide), with two settings. The first uses the RIQE settings. The second uses the parameters
published for NIQE~\cite{mittal2013niqe}: $P = 96$, $C = 1$, $p = 0.75$, the whole image and
no masks. On a $512\times512$ CT slice this gives the 25 non-overlapping patches of a
$5\times5$ grid; patches with zero MSCN variance (constant air or padding) are left out, and a
score needs at least two defined patches rather than the 72 required by RIQE. The image
covariance estimated from 25 patches is singular, but its average with the full-rank
reference covariance in Eq.~\eqref{eq:score} is not. Under this setting every one of the
18{,}840 test images and of the 1000 LDCTIQAC images has a score. The reference NIQE model
fitted on the LIVE photographs is not used, so this baseline is a refit with NIQE parameters,
not NIQE itself. As a check of the code on its own domain, both photographic models are
refitted on half of the photographs, and the other half is degraded with white Gaussian noise
($\sigma$ = 5, 10 and 20 code values) and Gaussian blur ($\sigma$ = 1, 2 and 4 pixels).

\subsection{Hyperparameter selection}

We searched 168 settings on the inner validation split:
\begin{align*}
P &\in \{16, 24, 32\}, \\
C &\in \{1, 0.5, 0.25, 0.1, 0.05, 0.025, 0.01\}, \\
p &\in \{0.05, 0.1, 0.2, 0.3, 0.4, 0.5, 0.75, 0.9\}.
\end{align*}
The criterion declared before running was lexicographic: zero cases in which a filtered
full-dose image scores better than the original; then monotonicity at every step of the noise
and blur ladders; then bootstrap stability. No setting satisfied the first condition, which
we now regard as too strong in any case: a full-dose image still contains noise, and a metric
may legitimately prefer a lightly filtered version. We therefore report that rate as a
descriptive endpoint, the \emph{filtered full-dose preference rate}, and test with inserted
lesions whether the preferred filtering removes signal.

Inspection of validation results showed that the declared criterion was blind to the
ordering of reduced-dose images, and that it selected a setting that ordered reduced-dose
abdomen correctly in 44.4\% of the 232 scoreable validation pairs. The criterion was revised
after seeing validation data and before touching the test set: first, reduced dose ranked
worse in at least 95\% of pairs in both chest and abdomen; then the lowest filtered-preference
rate; then detection of relative noise; then stability. We report the test performance of
both choices. We release a single model rather than one per protocol, a deliberate choice for
ease of use whose consequences are reported in Section~\ref{sec:strat}.

\subsection{Divergence between models}

The divergence between two models uses the functional form of Eq.~\eqref{eq:score}, so that it
reads in score units. For a contrast between two groups of patients, the null distribution is
obtained by 200 random patient-level re-partitions of the pooled groups, which matches sample
sizes by construction. As a scale with physical meaning we use the divergence between models
fitted on the full-dose slices and on the reduced-dose reconstructions of the same slices, for
the 80 Siemens fitting patients, and report
\begin{equation}
\eta = \frac{D_{\text{obs}} - \operatorname{med}(\text{null})}
{D_{\text{dose}} - \operatorname{med}(\text{null}_{\text{dose}})},
\end{equation}
the excess over sampling noise in units of the excess produced by the simulated dose reduction.

\subsection{Statistics}
\label{sec:stats}

On the TCIA data, patients are the unit of the split and of every interval. Slices, dose
pairs, degraded and filtered versions of one patient are not independent, so 95\% confidence
intervals are percentile bootstraps over patients ($B = 2000$): each resample draws test
patients with replacement and keeps all their images together. Where every patient is at
100\%, that interval collapses to a point and expresses no uncertainty; we then also report
the number of patients in which every pair is correct, with an exact binomial interval.
Sign tests, where used, count patients rather than slices. Stratification contrasts use the
permutation null above.

LDCTIQAC does not distribute patient or source-slice identifiers, so its analysis is
exploratory. We infer the source slice of each image: images whose low-pass filtered versions
correlate above 0.995 are linked, and the connected components give 69 groups. Every group
contains a multiple of ten images, consistent with ten degraded versions per source slice and
with no source slice split between groups, but some groups merge adjacent slices. Intervals
resample these groups; we also report the median correlation among the images of one group
and the paired difference between the correlations of two models, resampled jointly, and
repeat the analysis for thresholds from 0.98 to 0.999 (32 to 93 groups). Clustering the
groups into seven clusters, as many as the patients, did not recover the patients: on visual
inspection several clusters mix different body habitus. No patient-level interval is
therefore available for LDCTIQAC.

A score change can be arbitrarily small, so preference rates are also given counting only
changes larger than $\epsilon = 0.01$, 0.05 and 0.1 score units, and changes caused by added
noise are compared with the noise floor. The score depends on the covariance of the image as
well as on its mean features. To see whether filtering lowers the score by moving the mean
towards the reference or by changing the covariance, we repeat the preference analyses with
$D_{\text{ref}} = \big((\nu_1 - \nu_2)^{\top}\Sigma_1^{+}(\nu_1 - \nu_2)\big)^{1/2}$, in which
the image covariance is dropped.

\section{Results}

\subsection{Constraints that CT imposes on NIQE}
\label{sec:constraints}

\begin{table}[tbp]
\centering
\caption{Denominators. On the TCIA data, patients are the unit of the split and of every confidence interval; LDCTIQAC patients cannot be identified, and its intervals resample inferred source-slice groups (Section~\ref{sec:stats}). Unscoreable: fewer than 72 valid patches in the image domain, the rule of RIQE; the baseline with the NIQE parameters uses its own rule (Section~\ref{sec:baselines}).}
\label{tab:flow}
\small
\begin{tabular}{>{\raggedright\arraybackslash}p{2.6cm}>{\raggedright\arraybackslash}p{3.3cm}>{\raggedright\arraybackslash}p{3.6cm}>{\raggedright\arraybackslash}p{1.9cm}>{\raggedright\arraybackslash}p{2.6cm}}
\toprule
Experiment & Available & Selection & Excluded & Analysed (patients / units) \\
\midrule
Model fitting & 158 fitting patients, 24 slices each & -- & -- & 158 / 3,792 slices \\
Hyperparameter search & inner split 118 / 40 patients & 4 evenly spaced slices per validation patient; all 24 for dose pairs & -- & 40 / 160 slices, 480 dose pairs \\
Reduced dose (test) & 20 Siemens patients with reduced-dose series & all kept slices & 0 unscoreable & 20 / 480 pairs \\
Noise, blur, filtering (test) & 40 patients, 960 kept slices & 6 evenly spaced slices per patient & 0 unscoreable & 40 / 240 slices, 7,200 filtered \\
Denoisers on reduced dose & 20 Siemens test patients & slices of the test bank with a reduced-dose counterpart & 0 unscoreable & 20 / 120 slices \\
Inserted lesions & 20 abdominal test patients & central slice; sites by rule & 2 without 4\,mm sites & 20 / 20 slices \\
Stratification & 158 fitting patients & per contrast & -- & anchor 80 patients \\
LDCTIQAC 2023 & 1000 images, 69 source-slice groups & all & 0 unscoreable & 1000 images \\
\bottomrule
\end{tabular}
\end{table}

\paragraph{Patch size is bounded by the slice.} On $512\times512$ slices with the body mask,
the mean number of patches inside the domain is 517 at $P = 16$, 222 at $P = 24$, 119 at
$P = 32$, 51 at $P = 48$ and 10 at $P = 96$. The $96\times96$ patches of the original work
leave too few patches to estimate a per-image covariance. At $P = 24$ every one of the
18{,}840 test images has a defined score; at $P = 32$, 658 (3.5\%) do not
(Table~\ref{tab:flow}).

\paragraph{The sharpness threshold is badly conditioned.} In CT the most active patch of an
image is a bone--air interface, with local $\sigma$ of hundreds of HU against about 15\HU in
parenchyma, so a threshold relative to the maximum keeps almost only edges: at the nominal
$p = 0.75$, 2--6 patches per slice survive. These two constraints are consistent with the
$16\times16$ and $p \approx 0.32$ reported in~\cite{gunawan2025}.

\subsection{A trade-off across the settings tested}

\begin{figure}[tbp]
\centering
\includegraphics[width=0.72\linewidth]{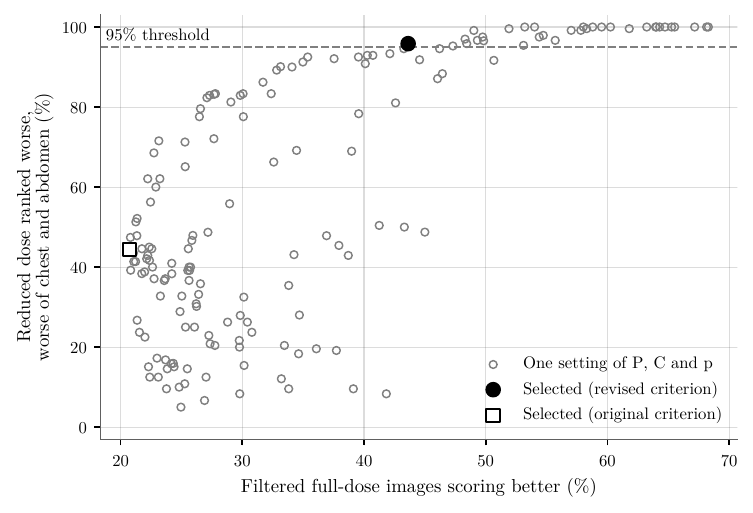}
\caption{Validation results for all 168 hyperparameter settings (inner split of the fitting
patients). Each open circle is one setting; the filled circle is the released model and the
square the setting chosen by the criterion declared before the search. Settings that rank
reduced dose correctly in at least 95\% of pairs in both regions prefer 43.6\% or more of the
filtered images; those with the lowest preference rates, about 21\%, rank fewer than half of
the abdominal pairs correctly.}
\label{fig:tradeoff}
\end{figure}

Figure~\ref{fig:tradeoff} shows the two criteria for all settings. Reduced-dose chest is ranked
correctly in every validation pair by every setting; abdomen separates them. Thirty-three
settings rank reduced dose correctly in at least 95\% of pairs in both regions; the best of
them prefers 43.6\% of filtered full-dose images to the original. Settings with the lowest
preference rate (about 21\%) rank reduced-dose abdomen correctly in fewer than half of the
pairs. The stabilising constant drives dose ordering: for $P = 16$ and $p = 0.2$,
reduced-dose abdomen is ranked correctly in 5.0\% of pairs with $C = 1$, 72\% with $C = 0.1$
and 83\% with $C = 0.01$. The selected model is $P = 24$, $C = 0.01$, $p = 0.5$, fitted on
121{,}213 patches from 3{,}792 slices. On the test set, the setting chosen by the original
criterion ranks reduced-dose abdomen correctly in 28.5\% [18.1, 40.9] of 214 scoreable
pairs (26 have no score at $P = 32$), that is, it prefers the reduced-dose image more than
two times out of three.

\subsection{Paired ranking of degraded images}

\begin{table}[tbp]
\centering
\caption{Paired ranking of degraded images on the held-out test split. Percentage of pairs in which the degraded image scores worse than its source (100\% is correct), with 95\% confidence intervals from a bootstrap over patients. Reduced dose: 240 chest and 240 abdominal pairs from 10 and 10 patients, reconstructions simulated by the data providers through projection-domain noise insertion. Added noise and blur: 240 full-dose slices from 40 patients (120 chest and 120 abdomen, 20 patients each). When every patient is at 100\%, the bootstrap interval collapses to [100, 100] and expresses no uncertainty; the second row gives instead the patients in which every pair is correct, with an exact binomial interval (for the rows at 100\% with 20 patients: 20/20, [83.2, 100.0]). That is the probability that all pairs of a patient are correct, not the rate of correct pairs.}
\label{tab:detection}
\small
\begin{tabular}{lcc}
\toprule
Degradation & Chest (\%) & Abdomen (\%) \\
\midrule
Reduced dose (10\%/25\% of routine), same slice & 100.0 {\scriptsize [100.0, 100.0]} & 95.8 {\scriptsize [91.2, 99.6]} \\
\quad patients with every pair correct & 10/10 {\scriptsize [69.2, 100.0]} & 6/10 {\scriptsize [26.2, 87.8]} \\
Added noise, +5\% of native & 85.0 {\scriptsize [75.8, 93.3]} & 89.2 {\scriptsize [81.7, 95.8]} \\
Added noise, +10\% of native & 92.5 {\scriptsize [85.0, 98.3]} & 98.3 {\scriptsize [95.8, 100.0]} \\
Added noise, +20\% of native & 97.5 {\scriptsize [95.0, 100.0]} & 100.0 {\scriptsize [100.0, 100.0]} \\
Added noise, +50\% of native & 100.0 {\scriptsize [100.0, 100.0]} & 100.0 {\scriptsize [100.0, 100.0]} \\
Added noise, +100\% of native & 100.0 {\scriptsize [100.0, 100.0]} & 100.0 {\scriptsize [100.0, 100.0]} \\
Gaussian blur, $\sigma=0.5$ px & 49.2 {\scriptsize [33.3, 64.2]} & 92.5 {\scriptsize [86.7, 97.5]} \\
Gaussian blur, $\sigma=1$ px & 99.2 {\scriptsize [97.5, 100.0]} & 100.0 {\scriptsize [100.0, 100.0]} \\
Gaussian blur, $\sigma=2$ px & 100.0 {\scriptsize [100.0, 100.0]} & 100.0 {\scriptsize [100.0, 100.0]} \\
\bottomrule
\end{tabular}
\end{table}

\begin{figure}[tbp]
\centering
\includegraphics[width=\linewidth]{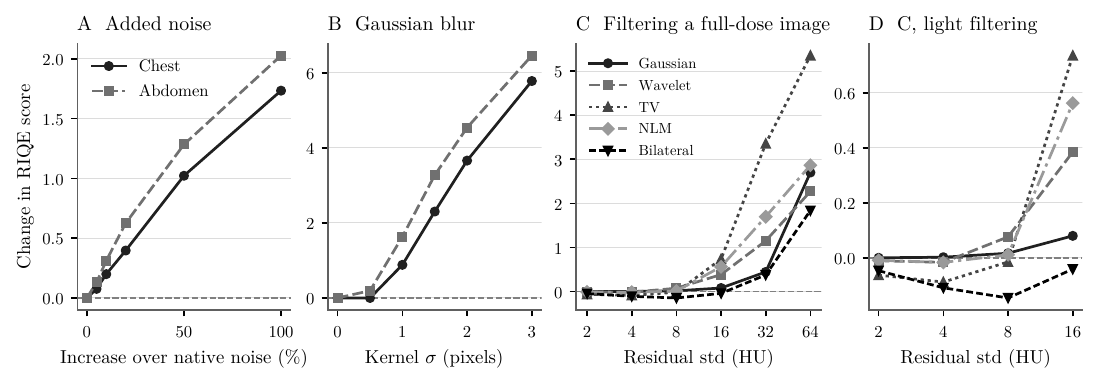}
\caption{Change in the RIQE score of the 240 held-out full-dose test images under controlled
degradation; positive values are worse. A: FBP-like noise added relative to the native noise
of each image. B: Gaussian blur. C: the five filters, calibrated per image to the same
residual standard deviation; values below the dashed line mean that the filtered image
scores better than the original. D: the lightest filtering levels of C on an expanded
scale. Points are medians over images; lines connect successive levels.}
\label{fig:response}
\end{figure}

Table~\ref{tab:detection} and Figure~\ref{fig:response}A--B report paired ranking on the 40
test patients. Simulated reduced-dose images score worse than the full-dose image of the
same slice in all 240 chest pairs and in 95.8\% [91.2, 99.6] of the 240 abdominal pairs, each
from 10 patients. Noise increases of 10\% or more are detected in 92.5--100\% of images, and
blur of one pixel or more in 99--100\%; half-pixel blur on thin sharp-kernel chest slices is
detected in only half of the images. Across the relative noise ladder, 86\% of images are
monotone at every step. Models fitted on bootstrap samples of the fitting patients differ
from the released one by a median of 0.12, against 1.84 between the full-dose and reduced-dose
models of the stratification anchor; a model fitted on 10 patients, the size of the corpus
in~\cite{gunawan2025}, differs by 0.66. Rankings produced by different bootstrap models agree
with a median Spearman correlation of 0.99.

\subsection{Preference for filtered full-dose images}

\begin{table}[tbp]
\centering
\caption{Preference for filtered full-dose images: percentage of cases in which the filtered image scores better than its unfiltered source (test split, 7200 filtered images from 40 patients), and in parentheses the median score change (positive = worse). A full-dose image still contains noise, so a preference is not by itself an error; whether the preferred filtering removes signal is tested in Table~\ref{tab:lesions}. Overall rate 41.2 {\scriptsize [34.8, 48.3]}\%; counting only changes larger than 0.01, 0.05 and 0.1 score units: 37.7, 27.8 and 19.2\%. Strength is the standard deviation of the residual inside the body.}
\label{tab:filtered}
\small
\begin{tabular}{lcccccc}
\toprule
& \multicolumn{6}{c}{Residual (HU)} \\
Denoiser & 2 & 4 & 8 & 16 & 32 & 64 \\
\midrule
Gaussian & 50 \,(+0.00) & 48 \,(+0.00) & 39 \,(+0.02) & 32 \,(+0.08) & 20 \,(+0.44) & 0 \,(+2.70) \\
Wavelet & 56 \,($-$0.01) & 56 \,($-$0.02) & 42 \,(+0.08) & 27 \,(+0.38) & 9 \,(+1.15) & 3 \,(+2.27) \\
TV & 82 \,($-$0.06) & 70 \,($-$0.09) & 50 \,($-$0.01) & 29 \,(+0.74) & 5 \,(+3.35) & 2 \,(+5.35) \\
NLM & 60 \,($-$0.01) & 57 \,($-$0.02) & 47 \,(+0.01) & 37 \,(+0.56) & 20 \,(+1.70) & 5 \,(+2.87) \\
Bilateral & 94 \,($-$0.05) & 96 \,($-$0.11) & 85 \,($-$0.15) & 59 \,($-$0.04) & 38 \,(+0.38) & 18 \,(+1.84) \\
\bottomrule
\end{tabular}
\end{table}

A full-dose image scores better after filtering in 41.2\% [34.8, 48.3] of 7{,}200 cases
(Table~\ref{tab:filtered}, Figure~\ref{fig:response}C--D). Many of these are near ties:
counting only improvements larger than 0.05 score units the rate is 27.8\%, and for Gaussian
filtering at the lightest levels the median change is zero. The preference is concentrated
on chest (61.0\% of cases, against 21.4\% on abdomen) and on edge-preserving filters: the
bilateral filter is preferred in 94--96\% of images at 2--4\HU and still in 37.5\% at 32\HU.
Dropping the image covariance from the distance does not remove it (45.5\% [39.0, 52.4]):
the mean features of the filtered images move towards the reference, and the preference
persists without the image covariance. On the simulated reduced-dose images the five filters at matched strength are
ranked only weakly consistently across images (median Kendall's $W = 0.23$ over the six
strengths).

Noise added at fixed levels is not preferred in the same way. On abdomen, added noise scores
better in 48\% of images at 1\HU, 37\% at 3\HU, 7\% at 10\HU and in none from 15\HU. On chest,
58--60\% of images score marginally better with 1 to 7.5\HU of added noise, with intervals
that include or touch 50\% (for example 59.2\% [46.6, 71.7] at 5\HU), and the median change,
$-0.001$ to $-0.004$, is a fraction of the noise floor of 0.013 measured at 5\HU.

\subsection{Preference and lesion signal}

\begin{table}[tbp]
\centering
\caption{Metric preference against lesion signal. Substrate: 20 held-out full-dose abdominal slices (18 with room for the 4\,mm lesions), one per patient, plus FBP-like noise at the level that separates full-dose from simulated quarter-dose abdominal images (local standard deviation 18.4\,HU, global 28.9\,HU); 64 noise realisations per condition. Each cell: percentage of images in which RIQE scores the filtered image better than the unfiltered one / median fraction of the matched-filter signal of a 4\,mm, +10\,HU lesion that survives the filter.}
\label{tab:lesions}
\small
\begin{tabular}{lcccccc}
\toprule
& \multicolumn{6}{c}{Residual (HU): preferred (\%) / signal retained} \\
Denoiser & 2 & 4 & 8 & 16 & 32 & 64 \\
\midrule
Gaussian & 0 / 0.98 & 0 / 0.97 & 0 / 0.93 & 0 / 0.86 & 0 / 0.70 & 0 / 0.34 \\
Wavelet & 0 / 0.99 & 6 / 0.98 & 39 / 0.96 & 61 / 0.89 & 0 / 0.69 & 0 / 0.29 \\
TV & 61 / 0.97 & 56 / 0.94 & 6 / 0.87 & 0 / 0.72 & 0 / 0.39 & 0 / 0.05 \\
NLM & 0 / 0.99 & 0 / 0.97 & 0 / 0.88 & 0 / 0.66 & 0 / 0.23 & 0 / 0.06 \\
Bilateral & 100 / 0.98 & 100 / 0.94 & 100 / 0.83 & 100 / 0.63 & 100 / 0.29 & 0 / 0.01 \\
\bottomrule
\end{tabular}
\end{table}

\begin{figure}[tbp]
\centering
\includegraphics[width=\linewidth]{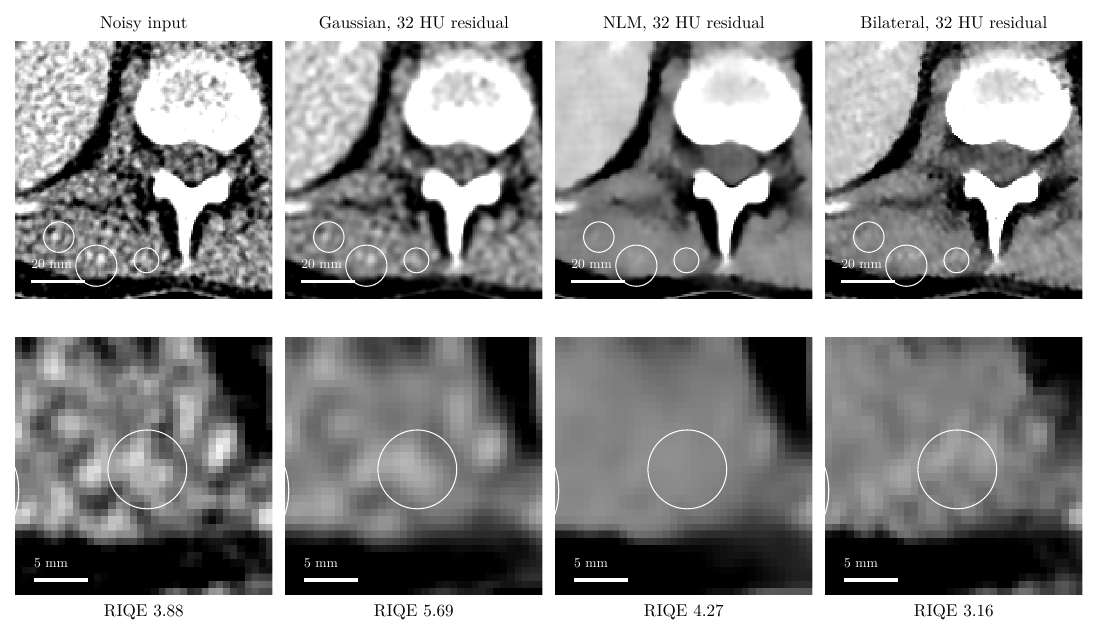}
\caption{One held-out abdominal slice with FBP-like noise at the level that separates
full-dose from simulated quarter-dose abdomen (local standard deviation 18.4\HU, global
28.9\HU) and three lesions inserted by the rule of the experiment (circles, from largest:
10\,mm $+25\HU$, 6\,mm $+15\HU$, 4\,mm $+10\HU$). Top: a region of the slice; bottom: a
24\,mm region centred on the 4\,mm lesion. The filters are
calibrated to the same 32\HU residual. Below each column, the RIQE score of the whole slice
without lesions, as in the experiment, lower meaning closer to the reference; the panels show
the slice with lesions. Display window $[-65, 185]\HU$; scale bars 20
and 5\,mm.}
\label{fig:examples}
\end{figure}

\begin{figure}[tbp]
\centering
\includegraphics[width=\linewidth]{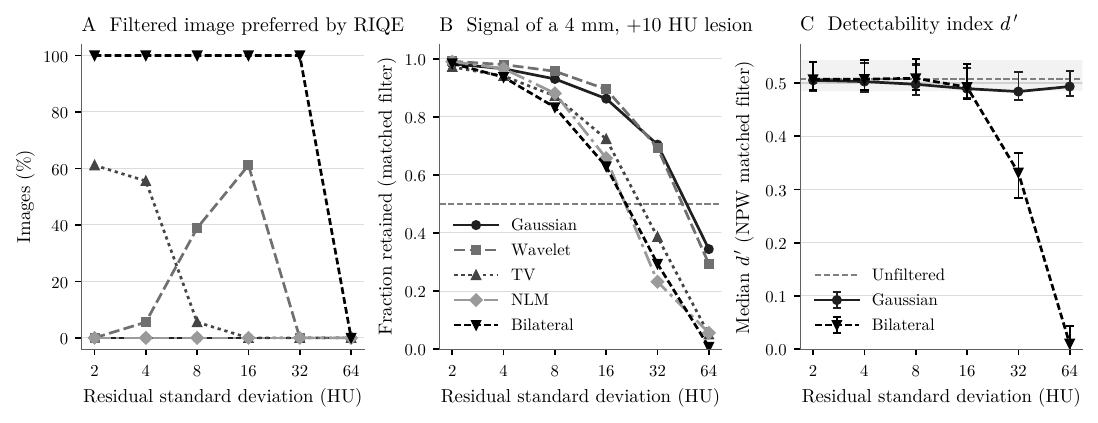}
\caption{A: percentage of the 18 held-out images in which RIQE scores the filtered image
better than the unfiltered one. B: median fraction of the matched-filter signal of a 4\,mm,
$+10\HU$ lesion that survives the filter; the dashed line marks half. C: median detectability
index $d'$ of the same lesion after Gaussian and bilateral filtering, the two filters compared
in the text, and without filtering (dashed line, shaded band); bars and band are 95\%
bootstrap intervals over patients. Up to 16\HU the two filters leave the same $d'$; at 32\HU,
where RIQE prefers the bilateral filter in every image, its $d'$ is lower.}
\label{fig:lesions}
\end{figure}

Table~\ref{tab:lesions} and Figures~\ref{fig:examples} and~\ref{fig:lesions} give the central
result. On the 18 abdominal slices with room for 4\,mm lesions, Gaussian smoothing and
non-local means are never preferred to the unfiltered noisy image, at any strength. The
bilateral filter is preferred in all 18 images up to a 32\HU residual, at which 29\% [25, 32]
of the matched-filter signal of the lesion remains, and in none at 64\HU. Total variation is
preferred in about 60\% of the images at the two lightest strengths, and wavelet shrinkage in
39--61\% at 8 and 16\HU, where it keeps about 90\% of the signal. Preference does not follow
the signal: at 32\HU, Gaussian smoothing keeps 70\% of the signal and non-local means 23\%, and
neither is preferred in any image.

The detectability index (Figure~\ref{fig:lesions}C) separates the filters less, since a
filter also changes the noise against which $d'$ is measured. Without filtering, the median $d'$ of the 4\,mm lesion is 0.51
[0.49, 0.54]. Up to 16\HU neither filter lowers it appreciably: 0.49 after both at 16\HU, a
paired difference of 0.005 [$-0.001$, 0.011]. At 32\HU it is 0.33 after bilateral and 0.48 after
Gaussian filtering, a paired difference of $-0.17$ [$-0.19$, $-0.14$]. At that strength RIQE
therefore prefers, in every image, the filter with the lower signal and the lower $d'$, and
never the one that keeps both; its preference for the bilateral filter ends between 32 and
64\HU, where the $d'$ of that filter collapses. The bilateral and Gaussian results are the
same when the image covariance is dropped from the distance. Averaging across filters hides
them: the median score over all filters worsens monotonically with strength.

\subsection{One model or several}
\label{sec:strat}

\begin{table}[tbp]
\centering
\caption{Divergence between sub-models fitted on two groups of patients (fitting split). $D$ uses the same functional form as the score. Null: median of 200 patient-level permutations of the pooled groups. $\eta$ expresses the excess over the null in units of the excess between models fitted on the full-dose slices and on the reduced-dose reconstructions of the same slices. All permutation $p = 0.005$ (the minimum attainable with 200 permutations). $^{\dagger}$Anatomy, kernel and slice thickness are not crossed in this collection; this contrast cannot serve as a clean anatomical control.}
\label{tab:strat}
\small
\begin{tabular}{lcccc}
\toprule
Contrast & Patients & $D$ & Null & $\eta$ \\
\midrule
\textit{Anchor: full vs reduced dose, same slices} & 80 / 80 & 1.84 & 0.25 & 1.00 \\
Abdomen 5\,mm: GE STANDARD vs Siemens B30f & 40 / 40 & 0.97 & 0.19 & 0.49 \\
Vendor: all GE vs all Siemens & 78 / 80 & 1.21 & 0.27 & 0.59 \\
Pixel-spacing tertiles, GE abdomen & 13 / 13 & 1.30 & 0.33 & 0.61 \\
Pixel-spacing tertiles, Siemens abdomen & 13 / 13 & 1.43 & 0.32 & 0.69 \\
GE: chest 1.25\,mm vs abdomen 5\,mm & 38 / 40 & 1.43 & 0.32 & 0.70 \\
Pixel-spacing tertiles, Siemens chest & 13 / 13 & 1.95 & 0.46 & 0.93 \\
Pixel-spacing tertiles, GE chest & 10 / 10 & 2.18 & 0.38 & 1.13 \\
Chest: GE STANDARD 1.25\,mm vs Siemens B50f 1.5\,mm & 38 / 40 & 2.36 & 0.32 & 1.28 \\
All chest vs all abdomen$^{\dagger}$ & 78 / 80 & 2.34 & 0.26 & 1.30 \\
Siemens: chest B50f 1.5\,mm vs abdomen B30f 5\,mm & 40 / 40 & 4.65 & 0.33 & 2.71 \\
\bottomrule
\end{tabular}
\end{table}

With the selected model, several contrasts diverge less than the simulated dose reduction
(Table~\ref{tab:strat}): vendor and kernel on abdomen ($\eta = 0.49$), vendor overall
(0.59), GE chest against GE abdomen (0.70), and pixel spacing within the abdominal cells
(0.61--0.69) and within Siemens chest (0.93). Chest contrasts diverge more: pixel spacing
within GE chest (1.13), vendor and kernel on chest ($\eta = 1.28$) and chest against abdomen
on Siemens (2.71). Since anatomy, kernel and thickness are not crossed in the collection, no
contrast can be attributed to a single factor. In validation, models fitted per protocol cell
ranked reduced-dose Siemens abdomen correctly in 100\% of pairs, against 95.8\% for the single
model with the same settings, but did not lower the preference for filtered images (46.0\%
against 43.6\% overall). They narrowed its spread across cells, from 17--76\% to 38--67\%: it
fell on Siemens chest (61.8\% to 37.7\%) and on the small GE CHST cell (75.8\% to 48.8\%), rose
slightly on GE STANDARD chest (61.6\% to 67.2\%) and rose on both abdominal cells (16.9\% to
43.3\% and 31.3\% to 39.5\%). A single model is a simplification with a measurable cost for
dose ordering; per-protocol models do not reduce the preference for filtering.

\subsection{Baselines and radiologists' scores}
\label{sec:ldctiqac}

\begin{table}[tbp]
\centering
\caption{RIQE against two models fitted with the same code on 119 CC0 natural photographs: one with the RIQE settings, one with the parameters published for NIQE ($P=96$, $C=1$, $p=0.75$, no masks). The reference NIQE model fitted on the LIVE photographs is not used. Rows 1--5: held-out test split. Rows 6--7: the LDCTIQAC~2023 training images, decoded as $\mathrm{HU} = 1400x - 1000$; 95\% intervals resample the inferred source slices; negative is the expected sign (lower score = closer to the reference). Within inferred group: rank correlation among the images of one group inferred from image similarity (the degraded versions of one source slice or of a few adjacent ones), median over groups. LDCTIQAC patients cannot be identified: these rows are exploratory.}
\label{tab:photo}
\small
\setlength{\tabcolsep}{3.5pt}
\begin{tabular}{lccc}
\toprule
& RIQE & Photographs, & Photographs, \\
& (CT) & RIQE settings & NIQE parameters \\
\midrule
Reduced dose ranked worse, chest (\%) & 100.0 & 100.0 & 98.3 \\
Reduced dose ranked worse, abdomen (\%) & 95.8 & 31.2 & 0.0 \\
Noise +20\% of native detected (\%) & 98.8 & 70.8 & 3.3 \\
Noise +100\% of native detected (\%) & 100.0 & 78.8 & 39.6 \\
Filtered full-dose preferred (\%) & 41.2 & 31.3 & 16.9 \\
LDCTIQAC: Spearman, pooled & $-$0.24 {\scriptsize [$-$0.32, $-$0.17]} & $-$0.17 {\scriptsize [$-$0.21, $-$0.12]} & +0.40 {\scriptsize [+0.31, +0.51]} \\
LDCTIQAC: within inferred group & $-$0.47 {\scriptsize [$-$0.54, $-$0.30]} & $-$0.19 {\scriptsize [$-$0.26, $-$0.08]} & +0.67 {\scriptsize [+0.63, +0.71]} \\
\bottomrule
\end{tabular}
\end{table}

\begin{figure}[tbp]
\centering
\includegraphics[width=\linewidth]{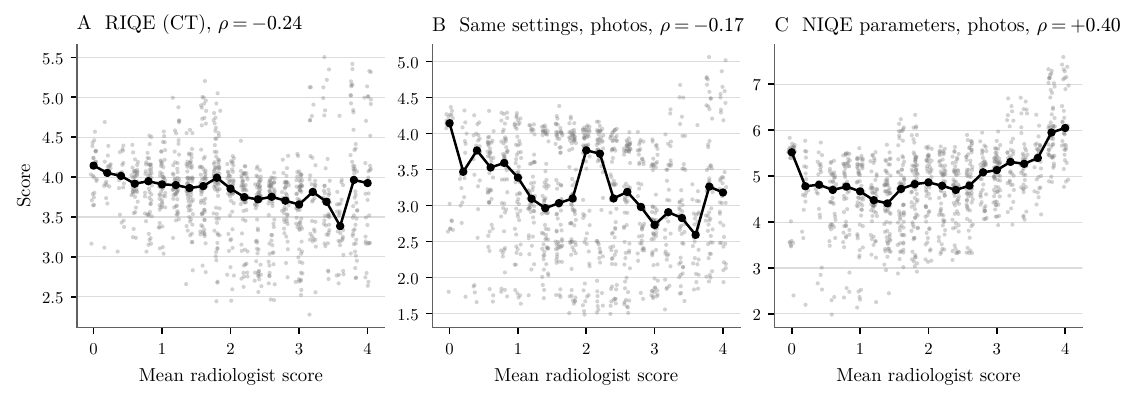}
\caption{Scores of the 1000 LDCTIQAC~2023 training images against the mean score of five
radiologists. A: RIQE. B: the same code fitted on CC0 photographs with the RIQE
settings. C: the same code fitted on the photographs with the parameters published for NIQE.
Grey points are images, jittered horizontally; black points are medians per radiologist
score. Lower scores mean closer to the reference, so the expected correlation is negative.}
\label{fig:ldctiqac}
\end{figure}

Fitted on 119 CC0 photographs with the RIQE settings, the same code ranks the test images
with a Spearman correlation of 0.84 to RIQE, but ranks reduced-dose abdomen correctly in only
31.2\% [20.8, 42.9] of pairs and detects a doubling of noise in 78.8\% of images
(Table~\ref{tab:photo}). With the parameters published for NIQE ($P = 96$, $C = 1$, $p = 0.75$,
whole image) the photographic model inverts: it ranks every reduced-dose abdominal image
better than its full-dose counterpart and detects a doubling of noise in 39.6\% of images.
In the check on held-out photographs, both photographic models, refitted on half of the
photographs, score the other half worse after white noise of 5 to 20 code values (80--97\% of images) and after blur of 1 to 4 pixels (97--100\%). The check
shows a plausible response to noise and blur on photographs; it does not establish
equivalence with the original NIQE implementation, and the behaviour of the original NIQE
model on these images is not tested. The lower preference of the photographic models for filtered CT images reflects a weaker or
inverted response to noise rather than a better behaviour.

The LDCTIQAC analysis (Figure~\ref{fig:ldctiqac}) is exploratory, since its patients cannot
be identified (Section~\ref{sec:stats}). Decoded as $\mathrm{HU} = 1400x - 1000$ from the
normalisation stated by the organisers~\cite{lee2025ldctiqa}, all 1000 images are scoreable.
RIQE correlates with the radiologists' mean score with $\rho = -0.24$ [$-0.32$, $-0.17$]
resampling the inferred source slices; across grouping thresholds from 0.98 to 0.999 the
interval ranges from [$-0.36$, $-0.09$] to [$-0.29$, $-0.19$]. Within one inferred group,
which contains the degraded versions of one source slice or of a few adjacent ones and is the
closest available analogue of the paired comparison for which RIQE is intended, the median
correlation is $-0.47$ [$-0.54$, $-0.30$] ($-0.42$ to $-0.50$ across thresholds). The
photographic model with the RIQE settings reaches $-0.17$ pooled and $-0.19$ within inferred
group; the paired difference within inferred group, $-0.12$ [$-0.24$, $-0.05$], favours RIQE
at every threshold ($-0.11$ to $-0.18$, all intervals excluding zero), while the pooled
difference, $-0.07$ [$-0.15$, $0.00$], is not resolved. These intervals are conditional on the
inferred grouping and do not account for the dependence between slices of one patient, so
they are exploratory evidence. The photographic model with NIQE parameters correlates with the
wrong sign ($+0.40$ pooled, $+0.67$ within inferred group),
consistently with its inverted response to noise on CT. The challenge report gives $-0.71$
for NIQE on the challenge test set~\cite{lee2025ldctiqa}, with the reference model we do not use and on other
images; we cannot say whether the reference photographs, the preprocessing or the images
account for the difference. The only patient identifiable in both LDCTIQAC and our corpus
belongs to our test split.

\section{Discussion}

\paragraph{Ranking is not selection.} RIQE orders a degraded image correctly against its
source when the change is known to be a loss: more noise, blur, simulated lower dose.
Selecting among filtered images is a different decision, because filtering removes noise and
signal at once, and there the score does not follow the signal. The difference is invisible
to evaluations that report average scores across methods: in our data the median score over
all filters worsens monotonically with strength, which would suggest a well-behaved metric.

\paragraph{A mechanism consistent with the data.} The score is a distance from the mean
statistics of the reference corpus, and moving an image's features towards that mean lowers
the score whatever the direction of the move. The corpus pools protocols whose native noise
differs by a factor of four, so for a given protocol the reference acts as a target rather
than an ideal. Several observations fit this reading: filtered chest images, from the noisier
protocols, are preferred far more often than abdominal ones (61\% against 21\%); per-protocol
models, which move the reference to each protocol, narrow the spread of the preference across
cells; the preference persists when the image covariance is dropped, so a shift of the mean
features suffices to produce it; and the stabilising constant, through which absolute noise
amplitude enters the features, drives dose ordering. Added noise does not mirror filtering,
however: abdomen penalises it and chest is nearly indifferent, so a single preferred noise
level does not describe the data and features other than noise amplitude must be involved.
We state the mechanism as consistent with the data, not as demonstrated.

\paragraph{Why edge-preserving filters.} With $p = 0.5$ the reference is built from the most
active patches, dominated by edges and structure. Edge-preserving filters keep that content
and flatten the low-activity texture in which noise, and small low-contrast lesions, reside;
linear smoothing degrades the edges as well and is penalised. This, too, is a hypothesis
consistent with the data.

\paragraph{Photographic references.} With the parameters published for NIQE, our code passes
the photographic check and inverts on CT; whether the original NIQE model does the same is
not tested here. A plausible reading is that at
$C = 1$ the smooth soft tissue of a CT slice, with a local contrast of a few code values,
produces MSCN statistics far from those of photographs, and added noise moves them closer.
Whatever the cause, a photographic reference can rank a noisier CT image as more natural,
the same kind of failure reported for a NIQE variant on ultrasound~\cite{rodrigues2025review}.

\paragraph{What RIQE is for.} RIQE is suited to paired ranking: comparing an acquisition or a
reconstruction with a degraded or processed version of the same anatomy, for quality
assurance or dose monitoring. We have not validated it as a detector on single images, which
would require thresholds, false-positive rates and generalisation across protocols. Under
the conditions tested it should not be the sole criterion to choose the strength of a
denoiser, to compare denoisers or to serve as an optimisation target, because its preferences
among filtered images do not follow lesion signal. Whether the same holds for deep-learning
denoisers, whose textures differ from those of classical filters, is untested.

\paragraph{Selection criteria.} A reasonable criterion, declared in advance, chose a setting
that ranks simulated reduced-dose abdomen backwards on unseen patients. Validation against
dose reductions, rather than synthetic noise alone, was what exposed it.

\paragraph{Scores are relative.} A RIQE score is a distance from the declared model, not an
absolute judgement of diagnostic quality, and is not comparable with NIQE values reported
elsewhere. The model does not know what a lesion is.

\section{Limitations}

No reader study was performed and no correlation with diagnostic performance is claimed; the
LDCTIQAC comparison measures agreement with perceived quality on abdominal images from seven
patients who cannot be identified, so its intervals rest on inferred groupings. The reduced-dose images are simulated by projection-domain noise insertion, not
acquired at lower dose, and exist only for one vendor. The corpus covers two vendors, five
protocol cells and two anatomical regions from one collection; head CT was excluded because
it remains under controlled access. Kernel, slice thickness and anatomy are not crossed, so
several contrasts identify combinations of factors. Motion and streak artefacts are not
detected directly. Reduced-dose test pairs come from 10 patients per region. Only five
classical filters and a finite grid of settings were tested; deep-learning denoisers, which
are the main use case, were not. Lesions are inserted rather than real, sit mostly in muscle
rather than in liver, and are evaluated by a matched filter that is not calibrated to human
readers, on 18 abdominal slices. The noise used for the relative ladders and the lesion
experiment is synthetic and approximates, rather than reproduces, the native noise power
spectrum. The features were implemented from the published description and were not compared
with the original implementation, and the original NIQE model was not used, by design; the
NIQE-parameter baseline is our refit. The hyperparameter criterion was revised after inspecting
validation data, before using the test set. All results depend on the declared intensity
mapping.

\section*{Data and code availability}

The code, fitting and evaluation scripts (MIT licence), the model and the corpus manifest
(CC~BY~4.0) are available at \url{https://github.com/Metallogik/RIQE}. The release used here,
v1.0, is archived on Zenodo with
\href{https://doi.org/10.5281/zenodo.23055559}{doi:10.5281/zenodo.23055559}. The experiment
outputs, tables and figures are regenerated from the public data by the scripts of the
release.

The model artefact records the intensity mapping, the hyperparameters and the criterion that selected
them, every slice used with its SHA-256, and the identity of the code. A verification script
refits the model from the public data and compares it with the released arrays. In the
recorded environment (Python 3.12.3, NumPy 2.5.3, SciPy 1.18.1, pydicom 3.0.2, OpenBLAS
0.3.34, x86-64, slices in manifest order) the refit is bit-identical;
with other versions or numerical backends the last digits may differ, and the script reports
the maximum relative deviation against a tolerance of $10^{-6}$.

\section*{Acknowledgements}

Data were obtained from The Cancer Imaging Archive (TCIA), collection LDCT-and-Projection-data,
funded by NIBIB grants EB017095 and EB017185 (PI: Cynthia McCollough). LDCTIQAC~2023 data
were obtained from Zenodo under CC~BY~4.0.


\end{document}